\documentstyle[prd,aps,preprint]{revtex}

\begin{document}
\draft

\preprint{Imperial/TP/94-95/38}


\title{Coarse-Grained Fluctuation Probabilities in the Standard
Model and Subcritical Bubbles}
\author{L. M. A. Bettencourt}
\address{The Blackett Laboratory, Imperial College,
London SW7 2BZ, U.K.}
\date{\today}
\maketitle

\begin{abstract}

We compute  systematically the probability for fluctuations of
the Higgs field, averaged over a given spatial scale, to exceed
a specified value, in the Standard Model.
For the particular case
of interest of averages over one coherence volume we show that,
even in the worst possible case of taking the one-loop improved
effective potential parameters, the probability for the field to fluctuate
from the symmetric to the asymmetric minimum before the latter becomes
stable is  very small for Higgs masses of the order of those of
the $W$ and $Z$ bosons, whereas the
converse is more likely. As such, metastability should be satisfied
dynamically at  the Electroweak phase transition and its dynamics
should therefore proceed by the usual mechanism of bubble nucleation
with subcritical fluctuations playing no particularly relevant role in it.
\end{abstract}



The possibility  of generating  the baryonic asymmetry of the
Universe at the Electroweak phase transition has attracted enormous
enthusiasm ever since it was realised by Kusmin, Rubakov and
Shaposhnikov \cite{KRS} that the Standard Model potentially satisfies
the three necessary Sakharov conditions \cite{Sakh}.
Whereas it is arguable whether the
$CP$ violating terms provide the necessary
magnitude for the observed
baryon to entropy ratio of the Universe \cite{FS},
in the minimal Standard Model,
the out of equilibrium bubble growth, characteristic of a  strongly
first order transition,
is certainly a {\it sine qua non} condition for nearly all electroweak
baryogenesis scenarios \cite{AllBubbles} to thrive.

The way of determining the actual order of the electroweak transition
is, in itself, a controversial matter. Arguments based on the shape of a
loop expanded effective potential about a homogeneous mean field background
constitute most of the evidence for a first order phase transition
\cite{DineL,Veff},
but in themselves, and mostly due to unavoidable infra-red
divergencies, fail to make it precise
how strong it can be, i.e.,
whether the phase transition dynamics is dominated by a few
thin-walled fast growing  bubbles.

Furthermore, the presence of important non-perturbative effects
seems undisputable and apart from making the perturbative
expansion meaningful in the infrared, was argued by Shaposhnikov
\cite{Non-pert} could render the phase transition more
strongly first order.
This picture seems to be supported by lattice simulations of the three
dimensional time-compactified effective theory\cite{Simul}.

The essential argument for the the transition to be weaker,
was suggested by Gleiser, Kolb
and Watkins \cite{GKW} and supported by Tetradis \cite{Tetra}.
They argued that
for sufficiently shallow barriers separating the two minima of
the effective
potential, it would be possible for the Higgs field to interpolate
between
them leading to approximate phase equilibrium and ousting thereby
bubble nucleation as the mechanism dominating the phase
transition dynamics. Instead,  thermal equilibrium would approximately
prevail throughout the transition making it impossible for the third
Sakharov condition for baryogenesis to be fulfilled.
Such fluctuations were later argued to be improbable by Dine et al.
\cite{DineL}, following estimates for the scalar field two-point
function (the variance of a Gaussian fluctuation distribution) computed at one
correlation volume, $v=O(m(T)^{-3})$. Subsequently Gelmini and Gleiser
\cite{GG} using a very different point of view based on modeling the
evolution
of the bubble population in both phases, rekindled  the issue by
finding considerable phase mixing for Higgs masses larger than about
$55 \ GeV$. Among other differences, inherent to quite distinct approaches,
the  nucleation rate or probability for a fluctuation, coherent over a
given volume, to exceed a given amplitude seems to be at the heart of
the standing discrepancies.
In what follows we compute the latter
quantity by following Hindmarsh and Rivers \cite{HR} in showing how
general fluctuating probabilities over given volumes
can be systematically computed
and how they  are essentially characterised by a hierarchy of
coarse-grained $n$-point correlation functions.

To achieve this we write the Higgs sector of the Standard Model,
after shifting the scalar
field about a homogeneous background scalar field $\phi$,
in the usual cartesian decomposition, in terms of two real scalar
fields, $\varphi_1$ and
$\varphi_2$, such that
\begin{equation}
\Phi = \phi + \varphi_1 +i \varphi_2
\end{equation}  as
\begin{equation}
{\cal L}_H = -V(\phi) + {\cal L}_0 + {\cal L}_I,
\end{equation}
\noindent where
\begin{eqnarray}
{\cal L}_0 &=&  {1 \over 2} (\partial_\mu \varphi_1)(\partial^\mu \varphi_1)
 -{1 \over 2} m_1^2 \varphi_1^2 \nonumber \\
&+& {1 \over 2} (\partial_\mu
\varphi_2) (\partial^\mu \varphi_2)  -{1 \over 2} m_2^2 \varphi_2^2
\end{eqnarray}
and
\begin{eqnarray}
{\cal L}_I &=& -\lambda \phi \varphi_1(\varphi_1^2 + \varphi_2^2)^2
- {\lambda \over 4} (\varphi_1^2 +\varphi_2^2)^2 \nonumber \\
&+& g A^\mu(\varphi_1
\partial_\mu \varphi_2^2 - \varphi_2 \partial_\mu \varphi_1^2)
 + g^2 A_\mu A^\mu \left[ \phi \varphi_1  + \right. \nonumber \\
&+& \left. {1 \over 2} (\varphi_1^2 +\varphi_2^2) \right]
+ {g_Y \over \sqrt{2}} \bar \psi (\varphi_1 + i \gamma_5 \varphi_2)\psi.
\end{eqnarray}

This is usually the departure point for any perturbative computation.
In the expressions above $A_\mu$ denotes a generic gauge field with $g$
being its coupling to the Higgs and analogously for $g_Y$,
relative to the fermionic chiral field $\psi$.
The scalar excitation $\varphi_1$ is the Higgs field whereas
$\varphi_2$ is the Goldstone mode, which becomes associated with the
gauge field and is not observable.

The proposal in \cite{GKW} can  be translated into noting that,
given that for Higgs masses
above the experimental lower bound the barrier
between the two minima of the effective potential is quite shallow,
before the asymmetric one becomes energetically favoured, there
would be a significant probability for coherent
fluctuations in $\varphi_1$ to transpose it, leading  to
approximate phase equilibrium.
In order to understand if this scenario is realised or excluded in the
Electroweak phase transition  we therefore need to be able to compute
systematically the probability associated with these processes,
i.e., that of the deviation from the minimum of the
potential with a given amplitude $\bar \varphi$ and over a given volume
$v$, $p({\varphi_1}_v> \bar \varphi)$.

To achieve this consider a scalar theory characterised by the
euclidean action $S_4[\varphi]$ and its partition function
\begin{equation}
Z = \int_B D \varphi e^{- S_4[\varphi]},
\end{equation}
where $B$ denotes the integral over periodic field configurations in
imaginary time, with period $\beta = {1 \over T}$, ($k_B =1$).
Then (5) can be rewritten as
\begin{equation}
Z = \int D \varphi e^{- \beta H[\varphi]},
\end{equation}
where $H[\varphi]$ is
\begin{equation}
H[\varphi] \simeq S_{\rm eff}[\varphi] -{\beta^2 \over 24}\int dx
\left(\delta S_{\rm eff} \over \delta \varphi (x) \right)^2
+O(\beta^4).
\end{equation}
Here $S_{\rm eff}[\varphi(x)]$ is the effective three-dimensional
action obtained from
$S_4$ after integration over all modes, but the light $n=0$ one, in a
Matsubara field expansion. Whenever the scalar field couples to other
fields their effects  are included through  the resulting
effective self-energies and couplings for the scalar field.
At high temperature and in the vicinity of
a minimum of the action $S_{\rm eff}$ all non-leading terms in the
expansion (7) can be neglected and  we can take $H[\varphi] \simeq
S_{\rm eff}[\varphi]$.

Further, we will be interested in computing field fluctuations
over a given spatial volume $v$, in which the field takes an
average value $\varphi_v$,
\begin{equation}
\varphi_v = {1 \over v} \int_{x \in v} dx \varphi(x).
\end{equation}
Let $p(\varphi_v \geq \bar \varphi)$ be the probability that the field
$\varphi_v$ be larger than a given value $\bar \varphi$, both larger than
zero. Then
\begin{equation}
p(\varphi_v \geq \bar \varphi) = {1 \over Z} \int_{\varphi_v \geq \bar
\varphi} D \varphi e^{-\beta H[\varphi]}.
\end{equation}
The Chebycheff inequality permits us to write an upper bound for the
probability, as
\begin{equation}
p(\varphi_v \geq \bar \varphi) \leq {1 \over Z}e^{-\beta v j \bar
\varphi} \int D \varphi e^{-\beta H[\varphi] + \beta \int dx J(x)
\varphi(x)},
\end{equation}
where $J(x) = j I(x)$ and $I(x)$ is a window function restricted to
the volume $v$.
Next we can define a three dimensional generator of connected Greens
functions $W[J]$, as
\begin{equation}
W[J] = \beta^{-1} \ln \left[ Z^{-1} \int D \varphi e^{-\beta
H[\varphi] + \beta \int dx J(x) \varphi(x)} \right]
\end{equation}
and an associated free energy $F[\varphi]$ through its Legendre
transform
\begin{equation}
F[  \varphi] = -W[J] + \int J \varphi \qquad {\rm with} \qquad \varphi =
{\delta W \over \delta J(x)}.
\end{equation}
Both these quantities are restricted to the volume $v$ by the
convolution of the usual source with the window function $I(x)$.
The bound on the probability then becomes minimised by
\begin{equation}
p(\varphi_v \geq \bar \varphi) \leq e^{\beta W[J] -\beta \int J\varphi
\vert_{J=jI}} = e^{-\beta F[\bar \varphi]},
\end{equation}
with $j$ chosen such that
\begin{equation}
\int J {\delta W \over \delta
J(x)}\vert_{J=jI} = \bar \varphi j v.
\end{equation}

In general a perturbative expansion for $W[J]$ can be
performed,  in powers of the interaction Lagrangian allowing us to
compute the Green's functions by comparison with
\begin{equation}
\beta W[J] = \sum_n {\beta^n \over n !} \int dx_1...dx_n
G^{(n)}(x_1,...,x_n) J(x_1)...J(x_n).
\end{equation}
Imposing (14) in order to minimise the probability bound yields the
polynomial equation
\begin{equation}
\sum_n {1 \over (n-1)!} (\beta v j)^n G_v^{(n)} = \beta v j \bar \varphi,
\end{equation}
\noindent where $G_v^{(n)}$ are the  connected
Greens functions, coarse-grained on the volume $v$, defined as
\begin{eqnarray}
G_v^{(n)} &=& {1 \over v^n} \int dx_1...dx_n G^{(n)}(x_1,...,x_n)
I(x_1)...I(x_n)\nonumber \\
&\equiv& <(\varphi_v)^n>.
\end{eqnarray}
The upper bound on the probability is then obtained by exponentiating
\begin{equation}
\sum_n {(\beta v j)^n \over n!} G_v^{(n)} - \beta v j \bar \varphi,
\end{equation}
evaluated at the solution for (14). Given a specific theory we are then
ready to compute $p$, to any desired accuracy in the scalar
effective couplings.

In order to illustrate the method
consider, first, the simplest case of a free scalar field
theory, with potential
$V(\varphi) ={1 \over 2} m^2 \varphi^2$.
In this case we need only compute the two-point coarse-grained Green's
function, the probability bound becomes exact and the fluctuations are
strictly Gaussian.
We can  proceed to obtain, for a window function $I(x)$ with Fourrier
transform  $\tilde I(k)$,
\begin{equation}
\beta v <\varphi^2_v> = {1 \over v} \int {d^3 k \over(2\pi)^3} {n(\omega)
\over \omega} \vert \tilde I(k) \vert^2,
\end{equation}
where $w = \sqrt{k^2 +m^2}$ and $n(\omega)$ is the Bose-Einstein distribution.
We will use the usual Gaussian window function given by
\begin{equation}
I(x) = \sqrt{2 \over 9 \pi}e^{-{x^2 \over 2 R_0^2}}\qquad {\rm and}
\qquad \tilde I(k) =  v e^{-{k^2 R_0^2 \over 2}},
\end{equation}
where $v$ is the spatial volume $v={4 \over 3}\pi R_0^3$.
Then, from our previous discussion we will typically be interested in
evaluating (18), for volumes of the order of the correlation volume for
the field, $ v={4 \over 3}\pi R_0^3$ with $R_0=O({1 \over m})$. We
parameterise the number of correlation volumes by writing
$R_0= {\alpha \over m}$, where $\alpha$ is a positive number. Taking
$\alpha$ to zero corresponds to no averaging and $\alpha =1$
corresponds to one spherical correlation volume.
The dependence of the coarse-grained two-point function on $m\over
T$, for several values of $\alpha$ is depicted in Fig. 1. It shows
that it is very sensitive to the coarse-graining scale, especially
for small masses. In this regime the dependence on the mass scale is
approximately linear. This can be extracted analytically as follows.
Using  (20) as the explicit form for the window function
and noting that the integrand in (19) is only substantially different
from zero for
$ \alpha^2{k^2 \over m^2} \leq 1$ which can only
realised at small momentum, we can  expand $\omega$,
using the dispersion relation, to obtain approximately
\begin{equation}
<\varphi_v^2> \simeq  \int_0^{m\over \alpha} {k^2 d k \over(2 \pi^2) }
{T \over m} {1\over m} = {1 \over 6 \pi^2 } { m T \over \alpha^3}.
\end{equation}
This illustrates how  the linear regime in $m$ arises
for small masses and shows that
the corresponding slope varies with the inverse coarse-graining scale
cubed, resulting in the  sharp differences observed in Fig.1. This will
be the regime of interest when we focus our analysis on the Standard Model
at its phase transition, as it was pointed out in \cite{Tetra},
where  important overall coefficients were ignored, however.
The importance of the constant $1 \over 2 \pi^2$ in  the
integral (21), in reducing the
spread  of the field, at one correlation volume,
was pointed out in \cite{DineL}, and constituted the basis of their
argument against a significant role played by subcritical fluctuations
in the Electroweak phase transition.

We are now ready to compute fluctuation probabilities in the
Standard Model.

In our previous analysis the computability of the
probability bound (13) resulted from the assumption of convexity
for $H[\varphi]$ and from our ability to approximate it by $S_3[\varphi]$.
This obviously holds  for
a free theory as well as for an interacting theory at temperature much
higher than the critical temperature, in its symmetric phase.
For gauge theories such as the Standard Model, close to their critical
temperature,
perturbation theory is only well defined around the minima of their
{\em perturbative}  effective potential\footnote{Modulo infra-red
divergencies in the unbroken phase. The issue of gauge and
parameterisation dependence
necessarily arises, as well, affecting both
the effective potential and our probability bound to the same extent.
To one loop these issues probably prove immaterial, though
\cite{KKK}.}. We can, nevertheless, still
construct our probability bound,  around each of those
minima and thereby estimate the probability for a fluctuation of a
given amplitude. If large amplitude fluctuations are then found to
be likely, say, compared to the distance to the maximum of the
effective potential separating the two minima then we will conclude
that the scenario proposed in \cite{GKW} is likely and unlikely or
impossible if the converse happens.

To do this we have to compute the two and higher-order
coarse-grained thermal Green's functions for the Higgs field.
These can be computed directly  from the Higgs effective scalar
theory (2) given
its effective mass and couplings resulting from its interaction with
the other fields in the
model. The former, in the usual limit of zero momentum, is given by
the square root of the second derivative of effective potential,
evaluated at the corresponding minimum, $m_{\rm min}^2=V_{\rm eff}
(\varphi_{\rm min})^{\prime \prime}$, while the latter is given ,
in the same limit, by its forth derivative.

Extensive calculations of the Electroweak effective potential
exist in the literature up to order $\lambda^2, g^4$ \cite{Veff}.
These results, however, for their length and variety of terms are  hard
to manipulate. The terms that make the two loop potential result
qualitatively different from the one-loop improved  result of
\cite{DineL}, are associated with logarithms of particle masses
over the temperature and are relatively simple to isolate.
Their total effect can be approximated  by using the analogous result for
the $SU(2)$ theory, where there is no distinction between the $W$ and $Z$
masses, $M$, which is given  by \cite{Dine},
\begin{equation}
\delta V_{\rm eff} = -{51 \over 32 \pi^2} {m_W^4 \over \sigma^2}
\phi^2 T^2 \ln(M_W/T).
\end{equation}
Two-loop contributions arising from diagrams including the top quark
are also important given its large mass. We include such corrections
again as computed in \cite{Dine}. In what follows we will
take the top quark mass  to be $m_{\rm top}=170 GeV$.

To order $\lambda$ we can, then, write the probability bound (13) as
\begin{equation}
p(\varphi \geq \bar \varphi) \leq e^{{\bar \varphi^2
\over 2 G^{(2)}_v} \left( 1 + { G^{(3)}_v \over {3 G^{(2)}_v}^2}\bar\varphi
+ { G^{(4)}_v \over {12 G^{(2)}}^3_v} \bar \varphi^2 \right)},
\end{equation}
\noindent where $ G^{(i)}_v $ is the coarse-grained $i^{\rm th}$
point-function. Fig. 2 a), b) and c) show the lowest order diagrams
involved in computing the two, three and four point functions
respectively, with the shaded blobs denoting averaging of the external
legs over the volume $v$. Corrections of order $\lambda$ to the two point
function are at most of a few percent and are unimportant for the following
discussion. Higher order coarse grained Green's functions arise at higher
powers of $\lambda$ and their contribution can safely be neglected.

The probability bound, written as (23), is essentially dictated by
its first term,
involving the two point function alone.
Figure 4  shows the dependence of the square root of the two-point
function  on the Higgs
mass as well as the distance from the local minimum at $\phi=0$ to the
nearest inflection point, at the temperature, $T_{\rm crit}$, when both
minima are
degenerate, computed for the 1-loop improved effective potential
parameters \footnote{At this Temperature the 1-loop improved
effective potential
is  symmetrical about each of the local minima so no
distinction needs to be drawn between them.}.
We see that there is a small  probability,
about $1.27 \% $, say,  for $m_{\rm Higgs}=70 GeV$, for the field to
attain the inflection point. This translates into approximately
$3 \times 10^{-3} \%$ for the fluctuations to reach the maximum
separating the two local minima, which is obviously a very small value.

More important  than the smallness of the this probability is
perhaps to understand what happens at temperatures at which the minima are
non-degenerate.
Fig. 5 a) and b) display the same quantities as in Fig.3, but at a higher
temperature, just below that at which the new minimum appears, for
fluctuations around the symmetric and asymmetric minimum, respectively.
We see that, corresponding to a significant difference in shape about
each of the minima, the fluctuation probabilities are now quite different.
While it is now slightly more difficult for a large fluctuation to occur
for the field located at $\phi=0$,  it quite likely for the converse
to happen at the asymmetric minimum \footnote{In the computations of
the probabilities the coherence lengths at each minimum were used. It
should be noted, however, that, close to $T_{\rm min}$ these two
length scales can be quite different so that a coherence length
fluctuation attaining the asymmetric minimum finds itself being of a
fraction of the coherence length in the new phase and therefore has a
much larger probability to fluctuate back.}.
The probabilities above become approximately $33 \%$ and $1.22 \%$,
respectively.
This differential in the fluctuation
probabilities naturally decreases from $T_{\rm min}$ to $T_{\rm
crit}$, but until then introduces a fundamental qualitative feature.
In a dynamical
situation where the field fluctuates about each minimum, any field
attaining
the asymmetric minimum is  brought back to $\phi=0$, while the
opposite process is not likely enough to replenish it.
Using two-loop effective potential parameters, in the way mentioned above,
only makes this statement stronger, as is shown in Fig. 6.
As an example, the probabilities of the field to attain the inflection
point and maximum are now of approximately $0.92 \%$, and
$7 \times 10^{-9} \%$ at $T_{\rm crit}$, respectively.

Finally, we compute the change to the probability bound resulting from the
three and four-point coarse-grained functions in (25).
The diagrams in Fig. 2 b) and c) involve similar integrals to the usual
setting-sun and basket-ball diagrams respectively\footnote{Note, however,
that now there is only one   vertex, so that the result
is down by a power of $\lambda$ and no symmetry factors are involved.},
now convolved with window functions restricting each leg to the same
volume $v$. We compute them in this way, using
the integrals for their leading contributions from
\cite{Diagrams}, where full account of resummation was taken
care of, in the absence of coarse-graining.
The resulting contribution from the four-point function to the exponent
of the probability bound is of order
$0.015 \%$,  and causes very little change to our previous discussion, for
small amplitude fluctuations.
The correction due to the three point function is more relevant but
because the effective
vertex is proportional to $\phi$ matters only for fluctuations at the
asymmetric minimum.
As can be seen in Fig. 6, it contributes to increase the probability
relative to the the values
considered in our previous discussion. As such, it makes fluctuations
from the asymmetric minimum to the symmetric one more likely than their
converse even at the critical temperature,
when the effective potential is locally approximately similar for both
minima.
Differences to the Gaussian probability values
occur mostly for fluctuations of the same size or
larger than the two point function and accounts for a difference of
less that $5 \%$. For larger fluctuations the effect of the four-point
function starts being noticeable  compensating for the
increase due to three-point one. In the case of large enough fluctuations for
the field to attain the nearest inflection point, at the critical
temperature, this discrepancy is a fraction of a percent and thus makes
very little practical difference.

{}From the ensemble of our calculations we can therefore safely conclude that
coherence length fluctuations of the Higgs field have a very small
probability of interpolating between the minima of the effective potential,
even in the worst case scenario of taking 1-loop improved parameters.
This should mirror the actual field dynamics provided calculations
over a  homogeneous mean field background and equilibrium apply.
In this scenario,
fluctuations of the Higgs field from the asymmetric minimum to zero
are always more likely to happen than their converse, at and above the
critical temperature. Together these two conditions guarantee that
metastability is preserved in the Electroweak phase transition, for Higgs
masses smaller  or of the order of the gauge bosons and therefore the
phase transition dynamics should proceed by the
usual process of bubble nucleation. Finally, we note that our
computations should
have a natural translation into the coefficients used in the kinetic
equation of \cite{GG}. It would be  interesting to check,
if in this context, their utilisation could lead to the
reconciliation of both pictures.

\section*{Acknowledgements}

I would like to thank R. J. Rivers for having suggested the problem to
me and for many extremely useful discussions. The present research
was supported by  'J.N.I.C.T., Programa Praxis XXI', under
contract BD 2243/92 RM.

\section*{Figure Captions}

Fig. 1 - The dependence of the coarse grained two-point function
on $M/T$ for 0.01,0.1, 0.5 and 1 correlation volumes.

Fig. 2 a) b) and c) - Diagrams involved in computing the coarse-grained
two, three and four-point functions, respectively, in the probability
bound (23), to lowest order. The shaded circles denote spatial
integration of the external legs over the same volume $v$.

Fig. 3 - The variation of the  square root of the coarse-grained two-point
function with the Higgs mass, for 0.5, 1 and 2 correlation volumes
as well as that of the distance between the minimum of the
effective potential and the nearest inflection point, at $T_{\rm crit}$.
All quantities are calculated with 1-loop improved effective potential
parameters.

Fig.4 a) - The same quantities as in Fig.3, computed at the symmetric
minimum and at a temperature just below that at which the asymmetric
minimum appears.

Fig. 4 b) - The  same as Fig. 4 a), at the asymmetric minimum.

Fig. 5 - The same quantities as in Fig. 3, for the two loop
effective potential parameters, at the asymmetric minimum.

Fig. 6 - The Gaussian Probability function (dotted line),
obtained by neglecting the
three and four-point functions in (23), and the full quantity
computed to order $\lambda$ (dashed line).

\end{document}